# Chirality-biased point defects dynamics on a disclination line in a nematic liquid crystal

Andrzej Żywociński^$*, Katarzyna Pawlak^, Robert Hołyst^&

and Patrick Oswald$

$Ecole Normale Supérieure de Lyon, Laboratoire de Physique,

46 Allée d'Italie, 69364 Lyon Cedex 07, France

^Institute of Physical Chemistry of the Polish Academy of Sciences,

Department III, Kasprzaka 44/52, 01-224 Warsaw, Poland

&Department of Mathematics and Natural Sciences – College of Science,

Cardinal Stefan Wyszyński University, Dewajtis 5, 01-815 Warsaw, Poland

*corresponding author. E-mail: zywot@ichf.edu.pl; fax: +48 22 343 3333; tel: +48 22 343 3247

**Abstract**

Chiral additives in the nematic liquid crystal can alter the dynamics of point defects moving on a disclination line. They exert a constant force on defects, leading to the bimodal distribution of distances between them at long times. The evolution of the system of defects in the presence of chiral additives provides a proof of the existence of repulsive forces between the defects at large distances. We find that addition of sufficient amount of chiral compound removes all point defects from the system. The process is studied in the system of 8CB (4-n-octyl-4'-cyanobiphenyl) doped with the chiral compound S811 (from Merck Co.) and in the computer simulations.

## I. Introduction

Disclination lines are very common defects in nematic liquid crystals. They usually form a characteristic *"thread-like texture"*, which is easily observable in the microscope when a droplet of nematic phase is deposited on a glass plate or introduced by capillarity between two glass plates. Unfortunately, this texture is too complicated to study the fine structure of defects as they are strongly curved and strongly interact with each other. In particular, their molecular configuration is difficult to determine in those conditions. For that reason, physicists looked for more controlled configurations. One classical way to force the formation of a single defect is to confine the nematic in a cylindrical capillary with homeotropic anchoring on internal wall[1-4]. In that way, the molecules orient perpendicularly to the surface of the capillary which leads to the formation of a wedge disclination line[5] of topological strength +1. A similar method with flat capillary tubes was used subsequently to control the formation of +1/2 disclination lines.[6]

All these experiments together with general topological considerations[7] showed that defects of integer strengths are topologically unstable, *i.e.* can disappear by a continuous reorientation of the molecules, whereas defects of half-integer strengths are stable, *i.e.* impossible to eliminate in a similar way.

This difference between defects of integer and half-integer strengths is essential and has many experimental consequences. One of them is that ±1 disclination lines have in usual conditions a non singular core, whereas ±1/2 disclination lines possess necessarily a singular core. This difference is flagrant in thread-like textures in which ±1 line defects form "thick threads" (whose diameter and contrast are not well defined) whereas ±1/2 lines form "thin threads" with a sharp core which strongly scatters the light. This topological difference also arises in cylindrical capillary. In particular, it is very well known that the +1 line which has non-singular core provided that the radius of the capillary is large enough (typically more than a few hundreds of Å). The reason is that the director can freely rotate to align parallel to the capillary axis in its center. This phenomenon has been nicely named "escape in the third dimension" by Meyer[3]. It leads to the suppression of the core singularity in defects of integer strengths, which is very favorable energetically. It has also been shown that defects of half-integer strengths can *partly* escape in the third dimension for similar energetic reasons, although their singular core subsists because it cannot disappear topologically (the anisotropy of the Frank elastic constants is responsible for these distortions as shown first by Dzyaloshinskii[8]). As a result, the director field around a disclination line is rarely planar, even in the case of defects of half-integer strengths.[9]

This complete or partial rotation of the director along the defect axis has another consequence. Indeed, the director may "escape" in two opposite directions. For that reason, the point defects separating segments of line escaping in two different directions can form along the line. These point defects were known for a long time in cylindrical capillaries and have been studied intensively.[2-5] In particular, they nucleate in a large number when the sample is "quenched" abruptly in the nematic phase either from the isotropic phase, or from the smectic A phase. In both cases point defects occur randomly along the disclination line, with a density strongly depending on the quench velocity. In a cylindrical capillary, point defects are characterized by a radial or a hyperbolic molecular orientation around their singular centers. For that reason they are called R-defect (radial or "hedgehog" defect) and H-defect (hyperbolic or "antihedgehog" defect), respectively. These defects are directly observable in the microscope if we use a capillary with a large radius[2-5] (several micrometers). But they were proven to form also in cylindrical micropores of radius $0.05 \ \mu m < R < 0.5 \ \mu m$ of porous media as the Vicor. This result was obtained by Crawford *et al.*[10,11] from NMR studies. These authors observed that defects nucleate after a rapid quench from the isotropic liquid and then annihilate by pairs of opposite signs when their mean distance is less than typically the pore diameter. On the other hand, they did not observe their complete disappearance, even after a long annealing of their samples.

The question that immediately arises from these experiments is how the point defects interact. This problem was studied intensively by theorists. Two different answers are given in the literature. According to Peroli and Virga[12-14] the defects of opposite signs attract at short distance and do not interact when their distance typically exceeds the capillary diameter (screening effect). On the other hand numerical simulations by Semenov[15] show that defects attract at short distance and repel at long distance. These two types of interaction may explain as well the experimental results of Crawford *et al.*[10,11] in porous media as they lead both to a metastable array of defects after annealing of the sample. Finally, the attraction at short distances between R- and H-defects was studied experimentally by Cladis and Brand[4] in a large-radius capillary. They observed that the motion of the two defects is asymmetric, the R-defect moving much faster than the H-one. This phenomenon has not yet been explained but is certainly due to some backflow effects neglected in the calculations up to now.[16]

Recently, an alternative method was proposed to study the nucleation and the annihilation processes of point defects on a $-1/2$ disclination line[17-20]. As shown in Fig. 1, such a line occurs spontaneously near the tip of the meniscus which forms at the edge of a homeotropic nematic sample. In this experiment, the sample, which consists of two parallel glass plates with inner surfaces forcing a homeotropic anchoring, is partly filled with liquid crystal by capillarity. This geometry is interesting because it allows direct observations in the microscope of the point defects and their movements.

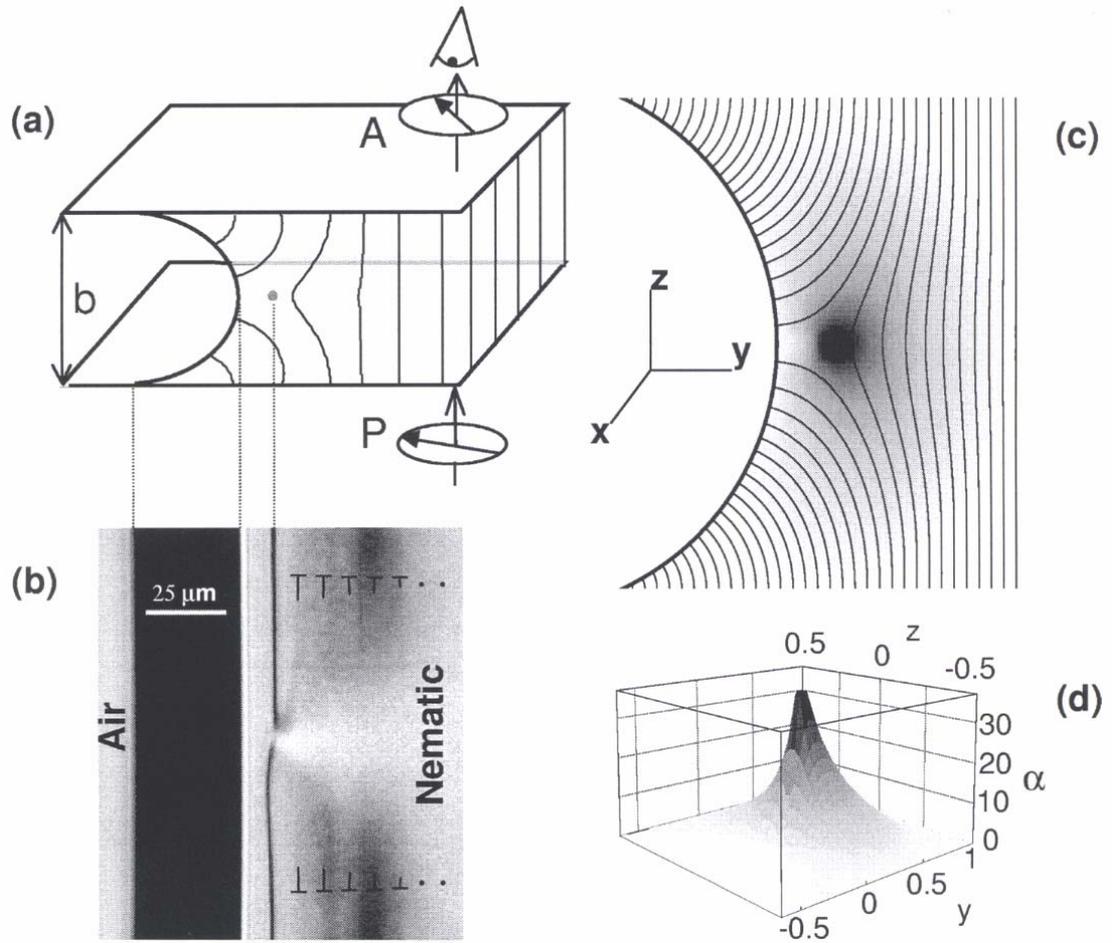

**Figure 1.** a) Formation of a -1/2 disclination line near the edge of a homeotropic nematic sample; b) top view of the sample when observed in the microscope with the polarizer (P) and the analyzer (A) partially crossed. The black region corresponds to the meniscus with the air. A point defect is visible on the disclination line, separating two zones tilted in opposed directions; c) calculated projection of the director field on the yz plane. The director is tangent to the solid lines. The intensity of the gray shading is an encoding for the magnitude of the escaped component of the director field along the line (x-component). The corresponding angle between the director and the yz plane is shown in (d) (from ref. [17]). Note that due to "escape in the third dimension", the director field is twisted, in particular between the line and the bulk sample. In addition, the twist (figured on the photo by the nails which represent the director field in the middle plane of the sample) changes sign on both sides of the point defects.

The main results obtained in this geometry are summarized as follows. First, the mechanisms of formation of the disclination line were analyzed both experimentally and theoretically.[19] It was shown that the line forms only if the anchoring of the molecules at the nematic-air interface is homeotropic (or sufficiently close to this orientation). Another observation was that the director field is not planar at equilibrium, but also escapes in the third dimension. In addition, the line can disappear (*i.e.* become virtual, which means that it is expelled from the nematic phase) when the anchoring penetration length K/W diverges (K is a Frank elastic constant and W the anchoring energy). The divergence leads to breaking of the homeotropic anchoring at the nematic-air interface. This phenomenon happens when approaching from a smectic-A phase because of the divergence of the elastic constant K (and more precisely of the bend and twist constants $K_2$ and $K_3$).[21] Similarly, the line can be expulsed from the sample by applying a strong enough electric field which favors the normal orientation of the molecules in the bulk sample.

Second, it has been observed that a great number of point defects can be nucleated along the line by the transient application of a strong enough electric field, or after an excursion into the smectic A phase. In both cases, the defects spontaneously nucleate when the line reappears in the sample. Note that the nucleation mechanisms of the point defects are still unknown in those conditions in spite of recent advances made in this field in directional melting.[20] It was then observed that the point defects collapse by pairs of opposite signs till forming a lattice of uniformly spaced defects. Finally, and this is the most important result for what follows, experimental studies of the time evolution of the defect distribution, coupled with numerical simulations, strongly suggest that defects attract at short distances and repel at large distances.

This repulsion at large distance between defects of opposite signs is counterintuitive, even if it seems quite well established in this special geometry. However, for R- and H-defects observed in the cylindrical geometry there is no definite answer concerning this point.

Faced with this confused situation, we addressed again this problem of the repulsion between defects and searched for a more direct way to prove its existence.

To reach this goal we doped our liquid crystal with a very small amount of a chiral component. Chiral molecules transform a nematic phase into a cholesteric one. To avoid complications, the concentration of chiral molecules was properly adjusted in order to obtain a pitch of the helix larger than the sample thickness. A direct consequence is that the director field remains homeotropic in the bulk sample (the cholesteric phase is unwound). On the other hand, right-hand and left-hand twist distortions are no longer equivalent in the vicinity of the disclination line because of the material chirality. A direct consequence is that two adjacent line segments separated by a point defect will have different energies per unit length as they involve twist deformations of opposite signs. The net result is that a constant force must act, for instance to the right, on all the defects of a given sign, whereas a force of opposite sign (but with equal magnitude) acts to the left on all the other defects of opposite sign. A crucial point is that these forces are independent of the distance between the defects. As a consequence, they will tend to make all the defects disappeared, whatever their distances, till all the line segments with the good twist merge in a single domain. This scenario should systematically be observed if there is no repulsion between defects, or if the "chiral force" is larger than the maximal repulsion force between defects of opposite signs, by admitting that it exists at large distances. On the other hand, the system should tend towards a metastable configuration consisting of an alternation of small and large domains (having, respectively, the "bad" and the "good" chirality), as long as the chiral force can be equilibrated by the repulsion force.

To test this idea, we performed the experiment and extended our previous numerical simulations by accounting for this constant additional force. Experimental results are presented in section II. The origin of the chirality-induced force on the defects is discussed in section III. The results of the simulations are given in section IV and conclusions are drawn in section V.

**II. Experiment**

As in previous experiments,[18-20,22] liquid crystal 8CB (4-n-octyl-4'-cyanobiphenyl, K24 from Merck Co.) was used. To induce chirality, 8CB was doped with a very small amount of the chiral compound S811 (from Merck Co.). Note that the concentration of S811 (0.15% by weight) was chosen in order to avoid the formation of cholesteric fingers in the bulk sample[21] (the cholesteric pitch is always larger than the sample thickness). The cell consists of two parallel rectangular glass plates treated for homeotropic anchoring with a solution of polyimide (3% mixture of compound #0626 in solvent #26 from Nissan Co.). The glass plates were successively spin-coated with a drop of a polyimide solution, dried under vacuum for 15 min at 80°C and polymerized during 45 min at 180°C. They were then glued together along their longer sides. Two 25-μm-in-diameter nickel wires served as spacers to fix the sample thickness. As explained in the introduction, the sample must be filled partially in order to obtain a meniscus between the nematic and the air. To realize this operation the cell was placed in an apparatus for directional solidification and filled in the presence of a large temperature gradient. The procedure, which is described in detail in ref. [17], allowed us to obtain a straight meniscus parallel to the longer side of the cell and perpendicular to the imposed temperature gradient (see Fig. 2).

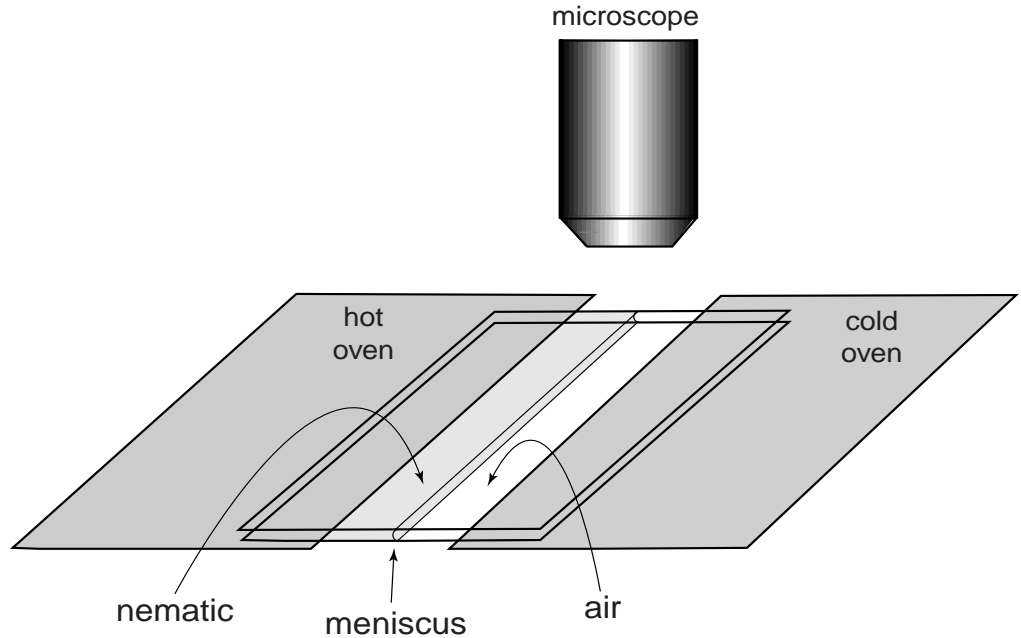

**Figure 2.** Experimental setup: cell in the directional solidification apparatus. The sample can be moved in the temperature gradient imposed by the two ovens. This way, the temperature of the meniscus can be easily and rapidly changed.

Point defects on the line are then prepared as follows. The sample was moved toward the cold side of the directional solidification apparatus till the temperature of the meniscus reached the temperature of the transition to a smectic A phase ($T_{NA} \approx 33.4°C$). During this passage the line disappears for the reasons explained in the introduction. The sample was then moved in the opposite direction till the meniscus reaches temperature $T_{NA} + \Delta T$ ($\Delta T$ values are given in Table 1). This temperature is known precisely as the position of the sample in the temperature gradient is measured within ±1 μm with an LVDT sensor (as for the temperature gradient, of the order of 10 K/cm, it is equally known within ±2%). Once the meniscus has reached its final temperature (*e.g.* $T=T_{NA}+0.32K$), the disclination line reappears while developing a great number of defects. At the beginning the defects rapidly annihilate by pairs of opposite signs. This process stops when the distance between the defect increases, what leads to the formation of a stable, but asymmetric, distribution of defects. The formation of the defects and their annihilation were recorded. A few pictures extracted from the movie are shown in Fig. 3.

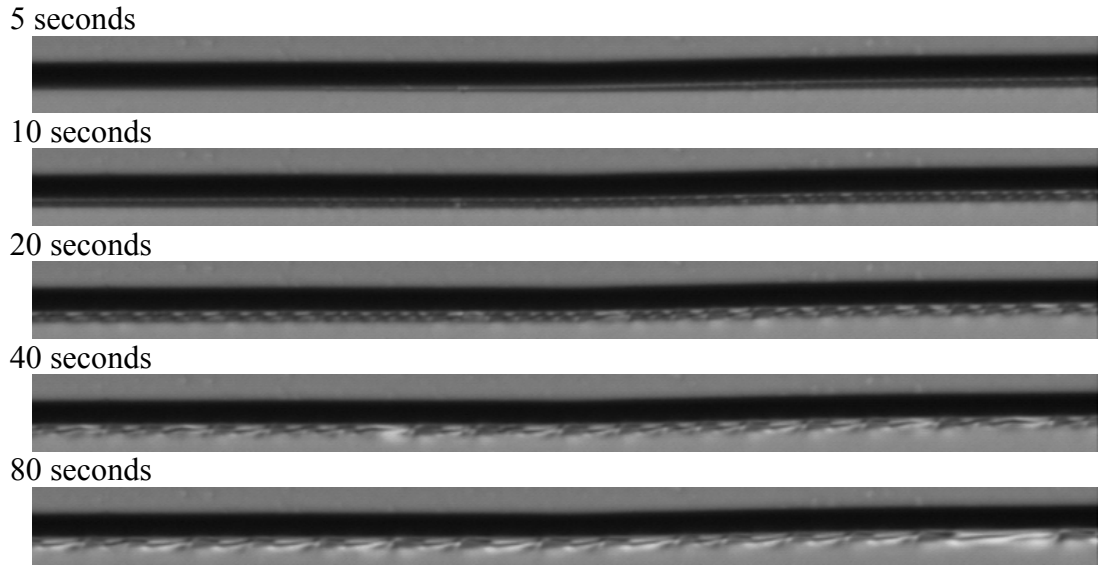

**Figure 3.** Time evolution of the point defects nucleated after a rapid passage from the smectic-A to the nematic phase; the black band corresponds to the nematic-air meniscus (air is at the top and nematic at the bottom); time is given on top of each image.

A quantitative analysis of the images showing the trajectories of the defects was performed using Igor-Pro Wavemetrics program. The images were analyzed and the positions of the defects were detected as a function of time. An example of such an analysis is presented in Fig. 4. The defects were detected using always the same algorithm with the same criteria of recognition. Note that at the beginning of the process many defects were not detected because of the too weak contrast of the images. In addition, defects separated by a distance smaller than 3 pixels were detected as one defect.

Similar experiments were performed at 5 different temperatures ranging between $T_{AN} + 0.32K$ and $T_{AN} + 1.93K$. The most stable distribution of the defects was obtained at T = $T_{AN} + 0.32K$ and is shown (after 4 hours of equilibration) in Fig. 5. In general, the defects distribution is very stable if there is no temperature gradient along the disclination line. The equilibrium distribution is then reached after 3 ÷ 4 minutes and remains stable during several hours (as shown in Fig. 5).

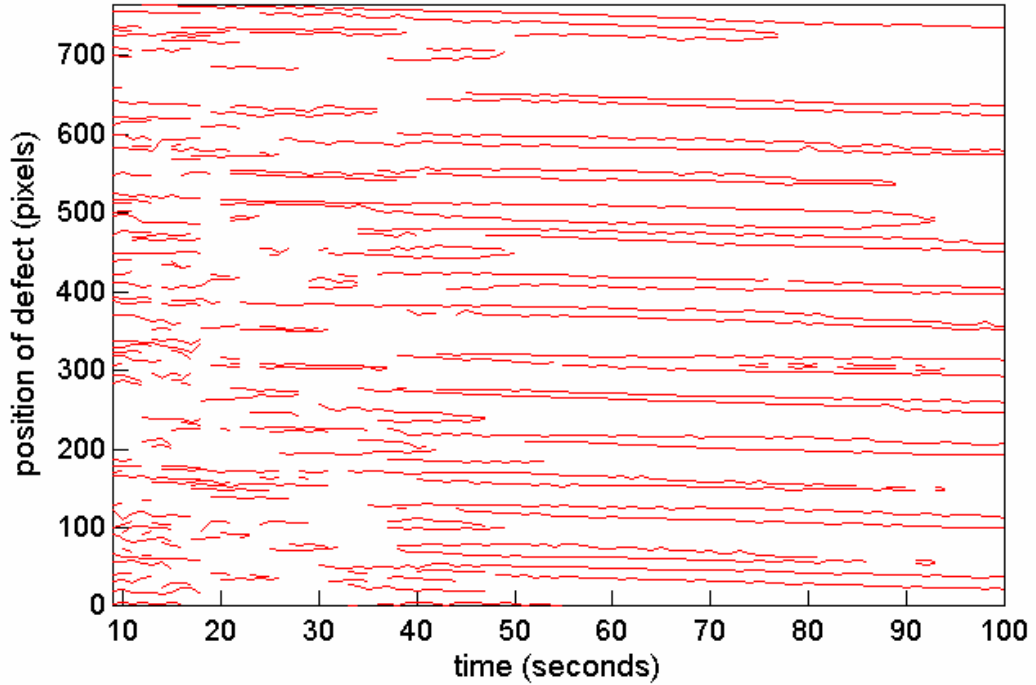

**Figure 4.** Results of the computerized analysis of the movie showing the early stage of evolution of point defects; 100 frames (whose a few examples are shown in Fig. 3) were acquired at frequency 1 frame/second.

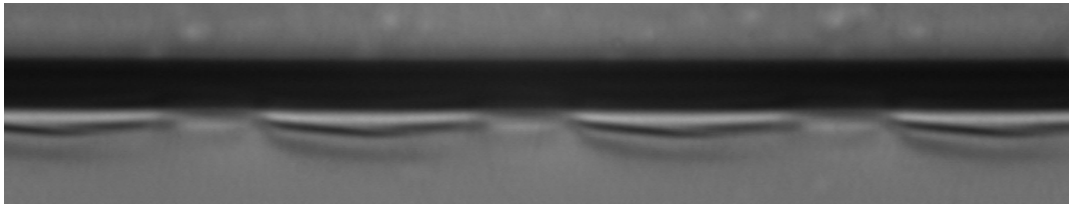

**Figure 5.** Equilibrium distribution of the defects at $T=T_{AN}+ 0.32K$; the image was taken 4 hours after the nucleation.

In some of the registered experiments the disclination line was not perfectly perpendicular to the temperature gradient. As a result the defects were very slowly drifting along the line, causing their disappearance at the end of the disclination line *i.e.* at the edge of the sample. This drift was already observed in previous experiment performed in directional solidification and was attributed to the fact that the point defect energy decreases when the temperature decreases (whence a constant drift of the defects towards the cold part of the sample).

The defects distribution and the drift velocity were found to depend on temperature. We did not performed systematic measurements so far but we observed that at temperature $T \geq T_{max} = T_{NA} + 1.54K$ all the defects rapidly disappear, even in the absence of a drift. This result shows that at temperature $T \geq T_{max}$, the repulsion force between defects can no longer equilibrate the "chirality-induced" force. In that case, all the defects collide and annihilate in a few minutes after their nucleation. The result is the formation of a straight line with the director escaping all along in the same direction (Fig. 6).

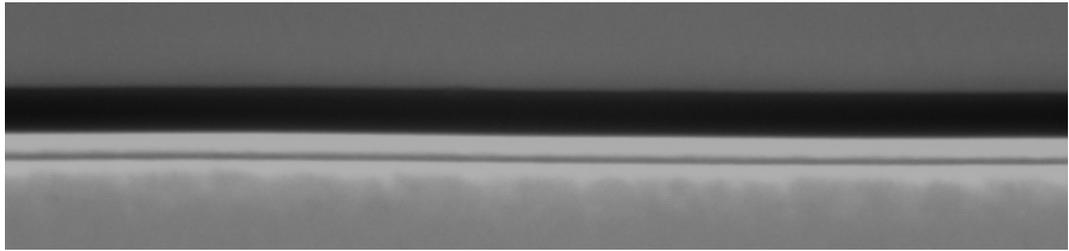

**Figure 6.** Disclination line after all the defects have disappeared. This final state is systematically reached at $T > T_{NA} + 1.54$ K, meaning that the "chiral" force always exceeds the repulsion force between defects.

The best recording of the evolution process was obtained at $T = T_{AN} + 0.54K$. Its analysis, starting 40 seconds after the nucleation, is shown in Fig. 7. One observes that during a relatively short period of time, four pairs of defects annihilate. It is worth noting that each annihilation gives more free space for the other defects, causing an increase of their separation distance. This observation is important as it is a direct proof for the existence of a repulsive interaction force between the defects.

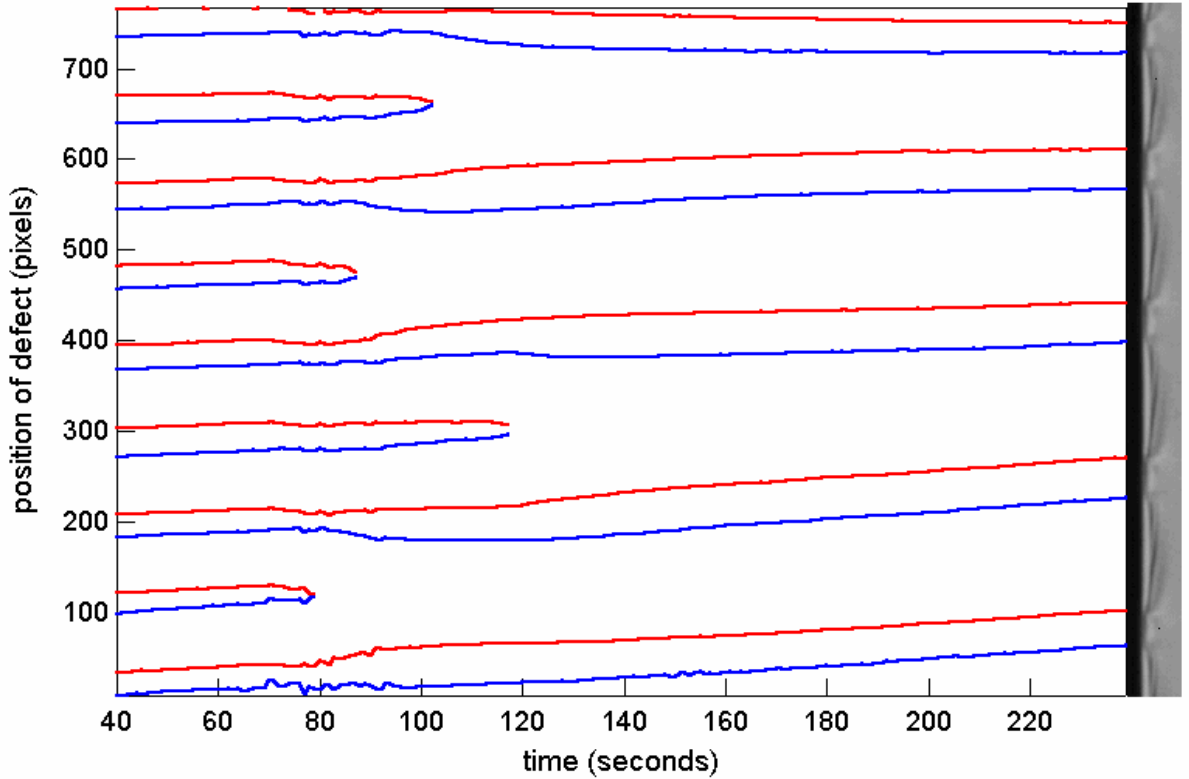

**Figure 7.** Position of the defects as a function of time starting 40 seconds after their nucleation at T=T$_{AN}$ + 0.54K (200 frames were analyzed); the process of annihilation of 4 pairs of the defects is shown; The inset on the right of the graph is a real image of the last analyzed frame.

To complete these observations, we measured the distance distribution over sequences of about 25 pairs of defects at different temperatures. The values of the ratio, $R = l/s$, of the averaged longer to shorter distances ($l$ and $s$, respectively) between the defects measured at five temperatures are given in Table 1. The increase of the ratio, $R$, which is a direct measurement of the asymmetry in the defects distribution, shows that the chiral force increases with respect to the repulsion forces as the temperature increases. This property will be qualitatively explained in section III.

**TABLE 1: Ratio, R = l/s, of the longer to shorter distances (l and s, respectively) between the defects as measured for different temperatures; temperatures are given as an increment over smectic-A – to – nematic phase transition, $T_{NA}$ = 33.43 K**

| ΔT (K) | R = l/s |
|--------|---------|
| 0.31 | 2.0 ± 0.4 |
| 0.32 | 2.4 ± 0.4 |
| 0.54 | 3.3 ± 0.6 |
| 0.99 | 4.7 ± 0.8 |
| 1.54 | 5.2 ± 1.6 |

It must be noted that the disclination line was generally not perfectly straight (this condition is very difficult to fulfill experimentally). For that reason, all the defects were not exactly at the same temperature what explains a variation of their distances (*ca.* 10%), the larger the higher was temperature. Additionally, the defects were often very slowly drifting. In spite of these artifacts, we observed that the distribution was always bimodal. An example of such a distribution measured 4 hours after the nucleation process is shown in Fig. 8. It is worth pointing out[18,20] that in absence of the chiral agent the distribution was unimodal with only one characteristic distance between the defects. In the next section the role of the chiral agent is explained and the chirality-induced force acting on each defect is estimated.

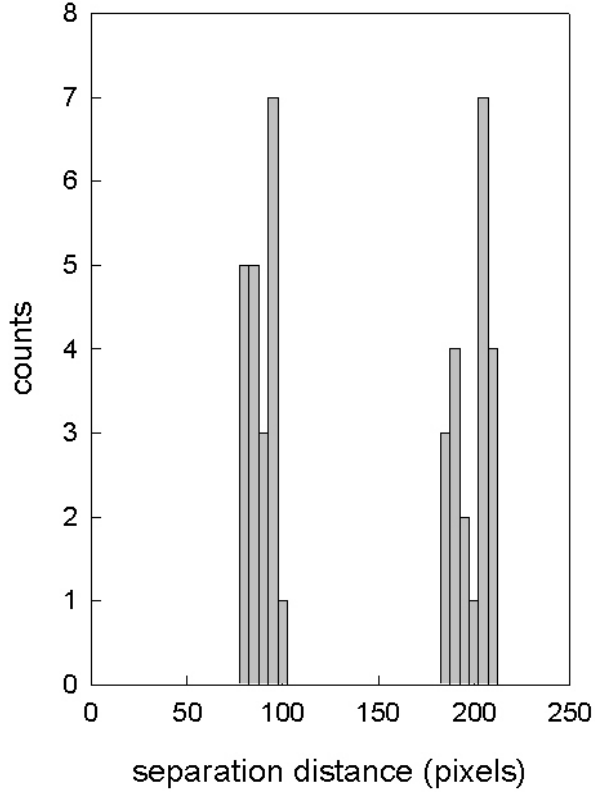

**Figure 8.** Histogram of the experimental distribution of the defects measured 4 hours after their nucleation at $T = T_{NA} + 0.32K$; the distribution is bimodal (two characteristic distances clearly emerge as can be seen in the photo of Fig. 7).

## III. Interactions induced by chiral additives

In the introduction, we have explained that the disclination line is not planar and tends to twist the director field (see fig. 1). This twist deformation is essentially localized between the line and the bulk sample. It typically extends over a surface area of size d × d. Let ±α be the tilt angle of the director close to the core of the disclination. The twist deformation energy per unit length of the line associated to this distortion is of the order of:

$$\frac{1}{2} K_2 \left(\frac{\alpha}{d}\right)^2 d \times d = \frac{1}{2} K_2 \alpha^2 \qquad (1)$$

where $K_2$ is the twist constant of the usual Frank elasticity.

In the absence of chiral additives, this distortion energy is obviously the same on both sides of any point defect in spite of the change of sign of the angle $\alpha$. For that reason, there is no force acting on the point defect (other than the forces exerted on it by the other defects).

Let us now consider the case of a cholesteric. In that case the material possesses a spontaneous twist $q_o$ proportional to the concentration of chiral dopant. As a result, the segment of a line situated on one side of the point defect where the tilt angle is $+\alpha$ must be distinguished from the segment of the line situated on the other side of the defect where the tilt angle is $-\alpha$. Indeed the distortion energy per unit length of a line becomes in the first segment

$$\frac{1}{2} K_2 \left( \frac{\alpha}{d} + q_o \right)^2 d^2 \approx \frac{1}{2} K_2 \alpha^2 + K_2 q_o d\alpha , \qquad (2)$$

whereas it is given by formula

$$\frac{1}{2} K_2 \left( \frac{\alpha}{d} - q_o \right)^2 d^2 \approx \frac{1}{2} K_2 \alpha^2 - K_2 q_o d\alpha \qquad (3)$$

in the other segment.
This energy difference is responsible for a net force acting on the defect of constant magnitude

$$f_{chiral} \approx 2 K_2 q_o d\alpha . \qquad (4)$$

This force acts on all the defects of the same sign (let us say +1). The same reasoning shows that defects of opposite sign (-1) experience a force of opposite sign but of the same magnitude.

It must be noted that close to the transition to the smectic phase $K_2$ diverges (it becomes infinite because twist distortion is forbidden in smectics) whereas $q_o$ tends to 0 (the cholesteric pitch diverges). On the other hand, the product $K_2 q_o$ remains constant as it only depends on the chiral power of the dopant (this is in fact the reason why $q_o$ tends to 0). As a consequence the chiral force acting on the defect is independent of temperature as a first approximation. The situation is completely different for the interaction force between two neighboring defects of opposite signs. Indeed, this force scales as the elastic constants $K_i$ (i=1,2,3) which all strongly increase[21] when approaching the transition to smectic-A. One thus expects that close enough to the transition, the chiral force can be equilibrated by the repulsion force, whereas at high temperature, the chiral force is always larger than the repulsion force. This qualitatively explains why stable patterns of defects form only close to the transition (at temperature $T<T_{max}$), whereas all the defects disappear when $T>T_{max}$.

In the following section we present a numerical simulation of the defect dynamics in the presence of a chiral agent.

**IV. Computer simulations of the dynamics of point defects**

We have studied the evolution of point defects distributed on the line interacting with a force $f(x)$ given by the formula[23]

$$f(x) = -(1-x)exp(-x), \qquad (5)$$

where $x>0$ is the distance between neighboring defects. This force is attractive at short distances and repulsive at longer distances.[23,24] We have assumed that the forces act between the nearest neighbors, but we have also verified that even if we assume that every defect interacts with every other defect on the line it does not change qualitatively and also quantitatively the results. Because the interactions between the point defects are related to the 3D distribution of the director field it is reasonable to assume that the interactions between the defects are strongly screened by the nearest neighbors and therefore this justifies the assumption of the nearest neighbors interactions. Our experiments also support correctness of such assumption, as can be seen from Fig. 7. Additionally, the chiral additives induce a constant force on the defects, as explained in the previous section. We assume that we have *N* defects of alternating type *(··+1-1··+1-1··+1-1··)* on the line of length *L*. The situation is schematically shown in Fig. 9.

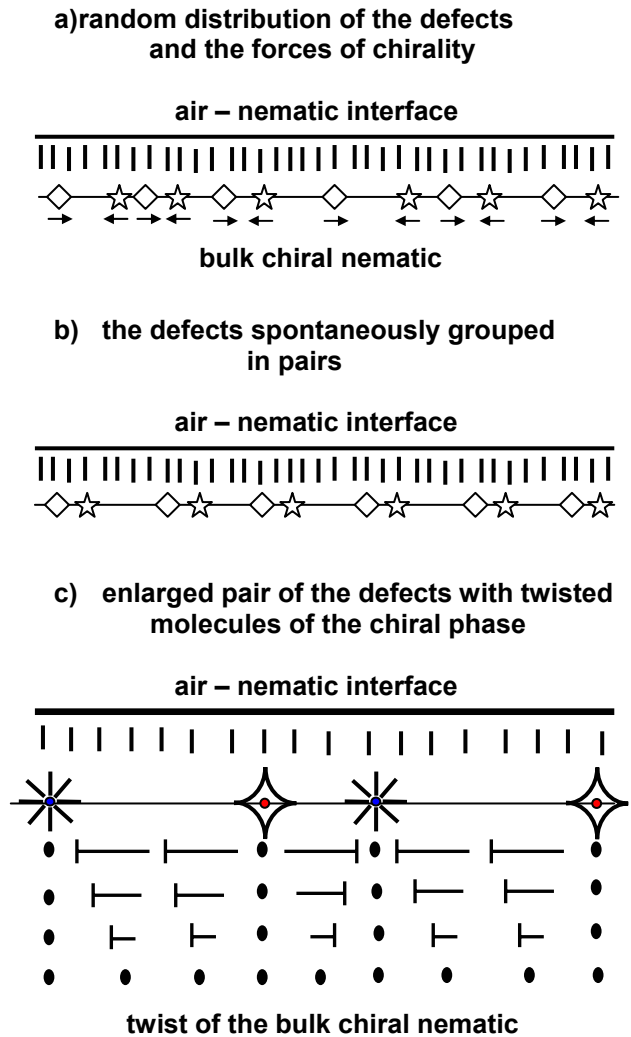

**Figure 9**. Schematic presentation of the defects self-assembling in pairs because of the chirality force (top view of the sample); star represents the radial defect (let's say "+") and rhomb represents the hyperbolic defect (consequently "-"). The director (represented by a nail in sketch "**c**") in the vicinity of the disclination line is, on the side of the bulk sample, alternately tilted to the top and to the bottom between each defect. On the other hand, it is perpendicular to the glass plates (and so represented by a dot) in each plane perpendicular to the line containing a point defect. Note that the director is similarly perpendicular to the glass plates far from the line in the bulk homeotropic nematic phase. For that reason, regions between the defects are alternately right-handed and left-handed twisted.

In the model we simply assume that the defects move under the influence of the force $f$ with a velocity proportional to it. When the friction coefficient is independent of both the distance between the moving defects and the type of the defect, we can write the following equations for the velocity $v_i$ of the *i-th* defect in the array (Stokes dynamics):

$$v_i = -f(x_{i,i+1}) + f(x_{i-1,i}) \pm f_{chiral}, \qquad (6)$$

Here, *f(x)* is the force acting between the defects (given by eq 5), attractive at short distances and repulsive at large distances, and $f_{chiral}$ (eq 4) is the constant force caused by chiral additives with "+" sign for one type of defects (say +1 defects) and "−" for the other (say −1 defects). As for velocity $v_i$, it equals

$$\frac{dy_i}{dt} = v_i. \qquad (7)$$

Here $x_{i,j} = |y_i - y_j|$ is the distance between the neighboring defects *i-th* and *j-th* and $y_i$ is the location of the *i-th* defect on the line.

In general all constants in the model such as viscosity, strength of the potential or its range rescale the time and the distance and therefore they do not explicitly appear in the model. It follows that all quantities in Eqs. (5÷7) are dimensionless. We took a reflecting boundary conditions with zero force acting on the defect at the boundaries and verified that the results do not change if we take as the boundary force a force produced by the mirror image of the last defect. In general the final results do not depend on the particular form of the interaction between the last (or first) defect on the line with the boundary.

As typical parameters we have used $L = 1000$ and $N = 1500$ with a random initial distribution of the defects on the line. Each defect (e.g. "+1") had an opposite defect ("-1") at each of its sides. Two defects were annihilated when the distance between them was less than 0.1. We used a simple Euler scheme to solve the equations with a time step of 0.1 and checked the results for the time step 0.01. We averaged the results typically over 400 runs *i.e.*

400 initial configurations. We have verified that the final results did not depend strongly on the initial distribution; even for an almost equidistant distribution of defects with small standard deviation (1 percent of the average distance) we have got the same results as with the random distribution. We have also noted that the final distribution of defects did not depend on the initial number of defects, $N$, on the line, providing that $N$ was large enough.

The dimensionless force, $f(x)$, given by eq (5) has a maximum of repulsion at the distance $x=2$ between the defects. At this distance it assumes a value $f(x=2)=0.135$. If the constant force $f_0$ exceeds the value -0.135, all the defects should annihilate since this force would overcome the repulsive barrier present in $f(x)$. Because $f_{chiral}$ is proportional to the concentration of chiral additives and presumably weakly dependent on temperature (see eq 4) we expect that at high temperature and/or high concentration of chiral additives all the defects should annihilate. This fact was observed in our experiments.

The histograms of the distances between the defects are shown in Fig. 10. For very small force, $f_{chiral}$, we find that the defects tend to form an equidistant array similarly as in the case for the nematic without the chiral additives. For larger $f_{chiral}$ the defects tend to group in pairs "$+1-1$". The distribution of distances is bimodal *i.e.* there is a shorter distance between the defects in pair and a longer distance between pairs. And this fact was also observed in our experiments.

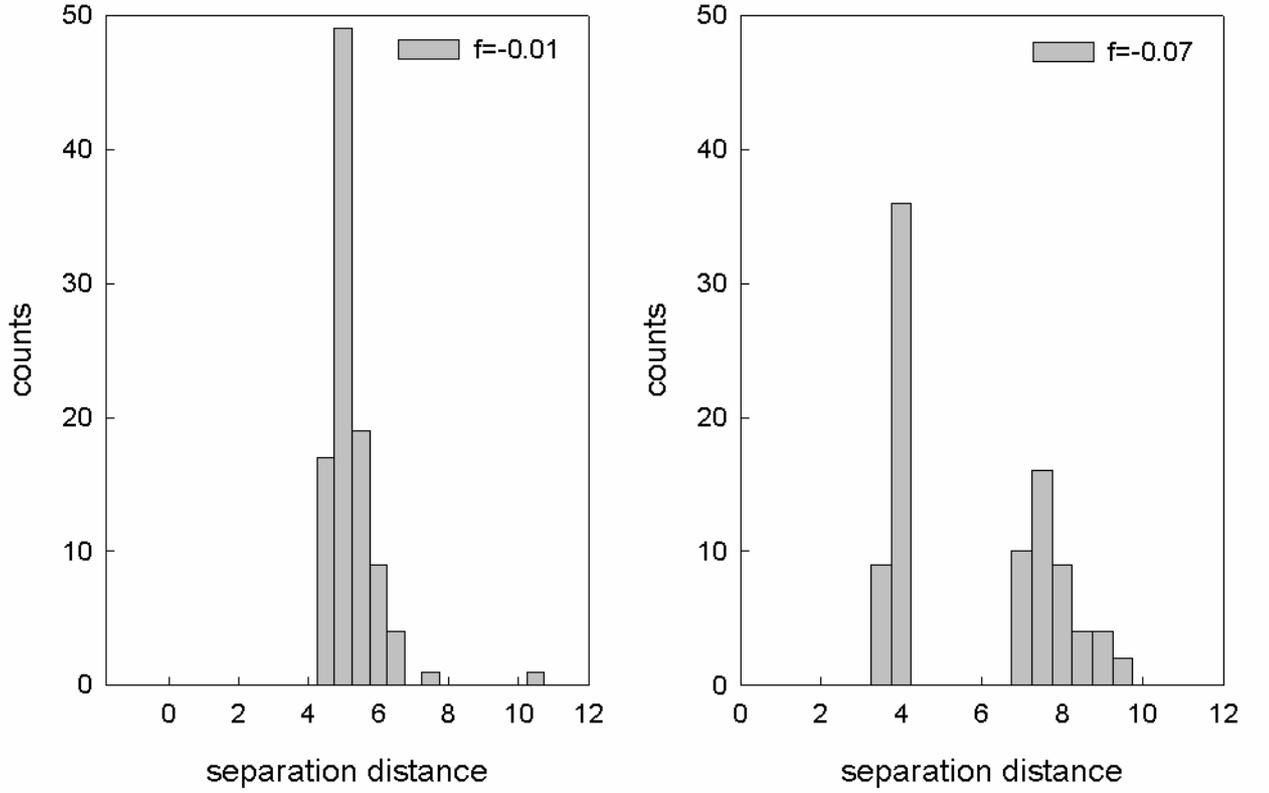

**Figure 10.** Histograms of simulated distribution of the defects for a constant chiral force: a) unimodal distribution for very weak force $f_{chiral}$ = -0.01 at low concentration of chiral additives; b) bimodal distribution for $f_{chiral}$ = -0.07 at higher concentration of chiral additives.

**V. Discussion and conclusions**

Our experimental and theoretical studies indicate that chiral additives allows the control of the distribution of defects on the line. Moreover, further experimental studies with chiral additives could help to determine in details the interaction potential between the defects. At present, we are certain that *"+1"* and *"−1"* defects attract each other at short distances and repel at large distances. Moreover, we have shown that the chiral additives can overcome the repulsive barrier and lead to the annihilation of all defects. More information will be obtained if we could predict the time and length scales from the microscopic calculations. So far microscopic calculations gave only conflicting results as to the nature of the interaction potential between the defects.[12,15,24] The dynamics of point defects is also more complex than described in the simulations. For example the mobility of defects (defined as the ratio between the force acting on the defect and its velocity) is certainly not a constant but must depend on the nature of the defects (because of probable backflow effects which should depend on the nature of the defect). We also expect that the mobility increases when the defects are very closely together as the distortions of the field in the bulk decrease in this limit. In the first approximation let us assume that the mobility is inversely proportional to the distance between the defects although a logarithmic dependence should be more appropriate. Even in this extreme case, we have verified that it does not change qualitatively the salient feature of the evolution, the main influence being to increase the time scale of the evolution. Another issue, not discussed in this paper in detail, concerns the core region of point defects. Its size might influence the interaction potential and dynamics of the defects. Finally we have not discussed the possible asymmetry of the dynamics of *+1* and *−1* defects on a −1/2 disclination line, although we observed that in our experiments. Such asymmetry exists in our geometry, as in cylindrical capillaries[4] between R- and H-defects, but is difficult to study in detail because of the presence of the temperature gradient which often causes a slow drift of the whole pattern. Further experiments at constant and homogeneous temperature are thus necessary to address seriously this point.

To conclude this article, let us emphasize that our experiment has a direct relation to one of the most important problems in condensed matter science. Namely, how to prepare a sample to achieve ordering at large distance while eliminating the defects. We have shown that, in our case, adding a very small amount of chiral impurities is sufficient to reach this objective.

**Acknowledgments**

This work was supported by the Committee of Scientific Research (Poland, KBN 2P03B00923 (2003÷2005), as well as the Polonium Program for bilateral cooperation between France and Poland. One of us (AŻ) would like to thank the people from École Normale Supérieure de Lyon for their hospitality.